\documentclass[twocolumn,aps,floats,letterpaper,floatfix, eqsecnum, groupedaddress]{revtex4-1}
\usepackage{graphicx}
\usepackage{dcolumn}
\usepackage{amsmath}
\usepackage{amsfonts}
\usepackage{amssymb}
\usepackage{epsfig,float,afterpage,wrapfig,psfrag}
\usepackage{natbib}

\newcommand{\beq}{\begin{equation}}
\newcommand{\eeq}{\end{equation}}
\newcommand{\bes}{\begin{subequations}}
\newcommand{\ees}{\end{subequations}}
\newcommand{\bea}{\begin{eqnarray}}
\newcommand{\eea}{\end{eqnarray}}
\newcommand{\ba}{\begin{array}}
\newcommand{\ea}{\end{array}}
\newcommand{\beqn}{\begin{eqnarray*}}
\newcommand{\eeqn}{\end{eqnarray*}}

\newcommand{\f}[2]{\frac{#1}{#2}}
\newcommand{\g}{\gamma}

\newcommand{\ra}{\rangle}

\newcommand{\dg}{\dagger}

\def\nn{\nonumber}

\newlength{\sizeonefig}
\newlength{\sizetwofig}
\begin{document}

\title{Two-photon scattering of a tightly focused weak light beam from a small atomic ensemble : an optical probe to detect atomic level structures}

\author{Dibyendu Roy} 
\affiliation{Theoretical Division and Center for Nonlinear Studies, Los Alamos National Laboratory, Los Alamos, New Mexico 87545, USA}

\begin{abstract}
We study two-photon scattering of a tightly focused weak light beam from a small atomic ensemble of two-level atoms (2LAs). This is similar to the scattering of photons from an atomic ensemble in a one-dimensional waveguide. The scaling of two-photon nonlinearity  at single-photon resonance shows a non-monotonic behaviour with an increasing number of few identical 2LAs. The two-photon nonlinearity decays monotonically with an increasing number of atoms for incident photons detuned from  single-photon resonance. Single-photon transport in two 2LAs is similar to that in a single $V$-type three-level atom (3LA). However two-photon transport in these two systems shows very different line-shapes. When single-photon transmission is zero in these systems, two transmitted photons are bunched together in a $V$-type 3LA, while their correlation is zero in two 2LAs. The difference in the two-photon line-shape persists for few 2LAs and 3LAs. Therefore, the two-photon scattering of a tightly focused weak light beam can be used as a probe to detect atomic level structures of different atoms with similar transition energies.    
\end{abstract}
\vspace{0.0cm}

\maketitle
\section{Introduction}
\label{intro}
A strong photon-photon interaction at the level of weak light field is one main challenge for realizing photonic quantum information devices. It is efficiently achieved by placing a nonlinear medium, such as an atom or a superconducting qubit in a cavity. The cavity greatly enhances the coupling between the medium and the photons. Recently a new approach to realize a strong two-photon nonlinearity at the level of few photons has been demonstrated using scattering of a tightly focused light beam from the dipole moment of an atom \cite{Gerhardt07, Zumofen08, Hwang09, Hetet11}. It can be realized easily by confining the light beam and the atom in a one-dimensional (1D) photonic waveguide \cite{Shen07, Chang07, Akimov07, Shi09, Roy10, Liao10, Longo10, Roy11, Zheng11}. Tight confinement of light fields in the waveguide directs the majority of the spontaneously emitted light from the atom into the guided modes, while local interactions at the atom induce a strong photon-photon correlation by preventing multiple occupancy of photons at the atom. 

A single two-level atom (2LA) is highly saturated by a single photon, thus it creates a strong optical nonlinearity for multiple incident photons. However an atomic ensemble can not be saturated by a single photon when the number of atoms is large, and the photon-photon interaction declines with an increasing number of 2LAs. We calculate a scaling of the two-photon nonlinearity with an increasing number of few identical 2LAs. The decay of two-photon nonlinearity shows a nonmonotonic behaviour with the increasing atom number at single-photon resonance of a 2LA. However the two-photon nonlinearity is diminished monotonically with increasing atoms for incident photons detuned from single-photon resonance. We also derive two-photon scattering by two different side-coupled 2LAs \cite{Rephaeli11, Zheng12}. The single-photon scattering by two different side-coupled 2LAs in a 1D waveguide is similar to that by a $V$-type three level atom (3LA) \cite{Witthaut10}. However, the nature of two-photon scattering in these two systems is quite different. When the single-photon transmission is zero in these systems, two reflected photons are anti-bunched in a $V$-type 3LA, but they are not anti-bunched in two 2LAs. Two transmitted photons at these parameters are bunched in a 3LA while there is no transmitted photon for the 2LAs. The difference in the two-photon line-shape persists for few 2LAs and $V$-type 3LAs. Therefore, we propose that the two-photon scattering line-shape of a tightly focused weak light beam can be used to probe efficiently atomic level structures of different type of atomic ensembles.    
\begin{figure}
\includegraphics[width=7cm]{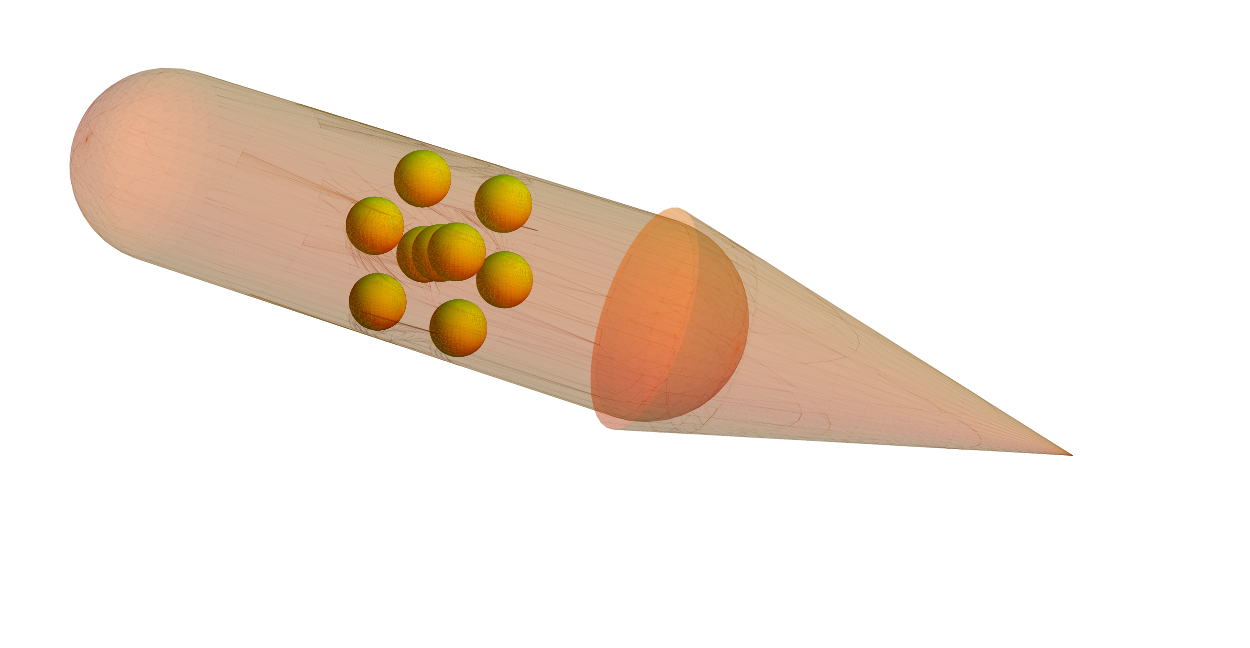}
\caption{A schematic of a small ensemble of atoms in a tightly focused weak light beam.}
\label{ensemble}
\end{figure}

The Hamiltonian describing the 1D waveguide-atom system also elucidates the interactions of a (tightly focused) light beam with an atom in three dimension (3D), when the beam is designed to mode-match the atom's radiation pattern \cite{Zumofen08, Hwang09, Gerhardt07, Hetet11}. Therefore, our present results are also relevant for the systems of Refs. \cite{Gerhardt07, Zumofen08, Hwang09, Hetet11} in 3D. The rest of the paper is organized as follows. In Sec.\ref{model} we introduce the general Hamiltonian.  In Sec.\ref{2tla} we calculate the single and two-photon dynamics for two different 2LAs in a 1D waveguide. The scaling of two-photon correlations with an increasing number of identical 2LAs is investigated in \ref{multi2la}. A comparison of two-photon correlations in few 2LAs and 3LAs is obtained in Sec.\ref{V3LA}.  We briefly discuss the effect of dipole-dipole interactions between atoms on photon-photon correlations in Sec.\ref{DipoleI}. In Sec.\ref{diss} we point out the advantages and limitations of the present technique and clarify some experimental issues. We include most of the technical details in the appendices.

\section{General model} 
\label{model}
We consider $N$ 2LAs side-coupled to the propagating left- and right-moving photon modes. These photon modes can be confined in a 1D waveguide \cite{Akimov07} or they can be created tightly focused by other experimental techniques \cite{Hwang09}. For simplicity of calculation, we assume all the atoms are connected to the chiral photon modes at a same position. This is a valid approximation for a small ensemble of atoms, $R<<\lambda$ where $R$ is radius of the ensemble and $\lambda$ is resonant radiation wavelength. This regime of interest is within the experimental reach in many atom-photon systems with microwave \cite{Wallraff04} and optical photons \cite{Akimov07,Faraon07,Lund08, Kim10} confined to a single-mode waveguide. Both real and artificial atoms can form the small ensemble of emitters. However, we are mainly interested in artificial atoms, such as superconducting qubits \cite{Wallraff04} and quantum dots \cite{Hogele04, Akimov07,Faraon07, Lund08, Kim10} ensemble in the solid-state environment. The advantages of these artificial atoms over the real atoms are their tunability of resonance frequency and a large dipole moment which helps to create strong light-matter interactions. The Hamiltonian of the full system of our interest,
\bea
&&\mathcal{H}=-iv_g\int dx ~[a^{\dagger}_{R}(x)\partial_xa_{R}(x)-a^{\dagger}_{L}(x)\partial_xa_{L}(x)] \nn \\&&+\sum_{l=1}^{N}\big(\tilde{\Omega}_la^{\dg}_{el}a_{el}+[\bar{V}(a^{\dg}_R(0)+a^{\dg}_L(0))\sigma_{l-}+H.c.]\big)~,\label{Ham}
\eea
where $\tilde{\Omega}_l=\Omega_l-i\g_l/2$. Here $v_g$ is the group velocity of photons, and $a^{\dagger}_{R}(x)~(a^{\dagger}_{L}(x))$ is a bosonic creation operator for a right (left)-moving photon at position $x$. We have set ground state energy of 2LAs to be zero, and $\Omega_l$ is energy of the excited state of $l$th 2LA. Here $a^{\dg}_{el}$  $(a^{\dg}_{gl})$ is a creation operator of the $l$th 2LA's excited (ground) state, and $\sigma_{l-}=a^{\dg}_{gl}a_{el}$. The coupling strength between the photon modes and a 2LA is $\bar{V}$. We are primarily interested in the regime where the atoms are strongly coupled to the guided modes of the waveguide. It implies that most part of the spontaneous emissions from the atoms is going into the guided modes which has been reported in experiments \cite{Lund08}.  Here we also incorporate spontaneous emissions from the atoms to the other non-guided modes which are not part of the left and right moving photon modes of the Hamiltonian in Eq.\ref{Ham}. This is done by including an imaginary term $-i\g_l/2$ in the energy of the excited atomic state within the quantum jump picture \cite{Carmichael93}. We can include a dipole-dipole interaction between atoms of the form $J\sum_{l,m=1}^{N}(\sigma_{l+}\sigma_{m-}+\sigma_{m-}\sigma_{l+})$ where $J$ is the interaction strength. It is straight forward to incorporate such interactions within our approach \cite{Roy12}. However we exclude such direct atom-atom interactions in our Hamiltonian to emphasize that the light-matter interactions create an effective photon-photon as well as atom-atom interactions. Later we mention how a direct dipole-dipole interaction term in the Hamiltonian can affect the photon-photon correlations. Next we transform the Hamiltonian using, $a_e^{\dg}(x)\equiv \f{1}{\sqrt{2}}(a^{\dagger}_{R}(x)+a^{\dagger}_{L}(-x))$ and $a_o^{\dg}(x)\equiv \f{1}{\sqrt{2}}(a^{\dagger}_{R}(x)-a^{\dagger}_{L}(-x))$ which decouple $\mathcal{H}$ in two parts, $\mathcal{H}=\mathcal{H}_e+\mathcal{H}_o$. One part $\mathcal{H}_e$ contains interaction of atoms with even photon modes while the other part $\mathcal{H}_o$ has only free propagating photons in odd modes. $\mathcal{H}_e=-i v_g\int dx ~a^{\dagger}_{e}(x)\partial_xa_{e}(x)+V(a_e^{\dg}(0)\sum_{l=1}^{N}\sigma_{l-}+H.c.)+\sum_{l=1}^{N}\tilde{\Omega}_la^{\dg}_{el}a_{el}$ and $\mathcal{H}_o=-i v_g\int dx ~a^{\dagger}_{o}(x)\partial_xa_{o}(x)$ where $V=\sqrt{2}\bar{V}$.  We set $v_g=1$. 
\begin{figure}
\includegraphics[width=7.5cm]{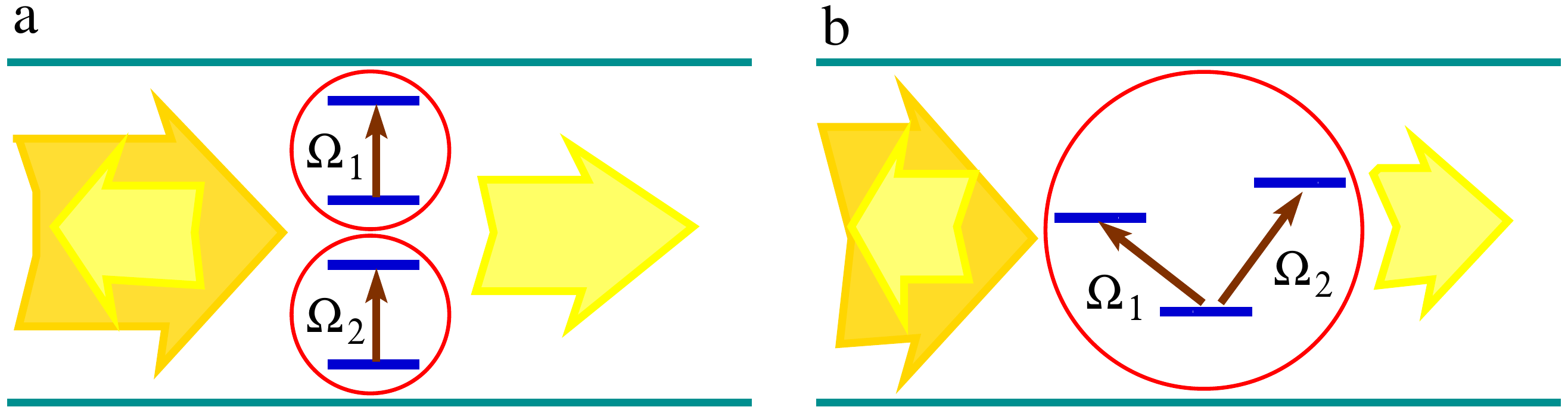}
\caption{(a) Two different two-level atoms and (b) a $V$-type three-level atom in a 1D photonic waveguide with incident, transmitted and reflected light beams.}
\label{3LA}
\end{figure}

\section{Two different two-level atoms}
\label{2tla}
\begin{figure}
\includegraphics[width=8.5cm]{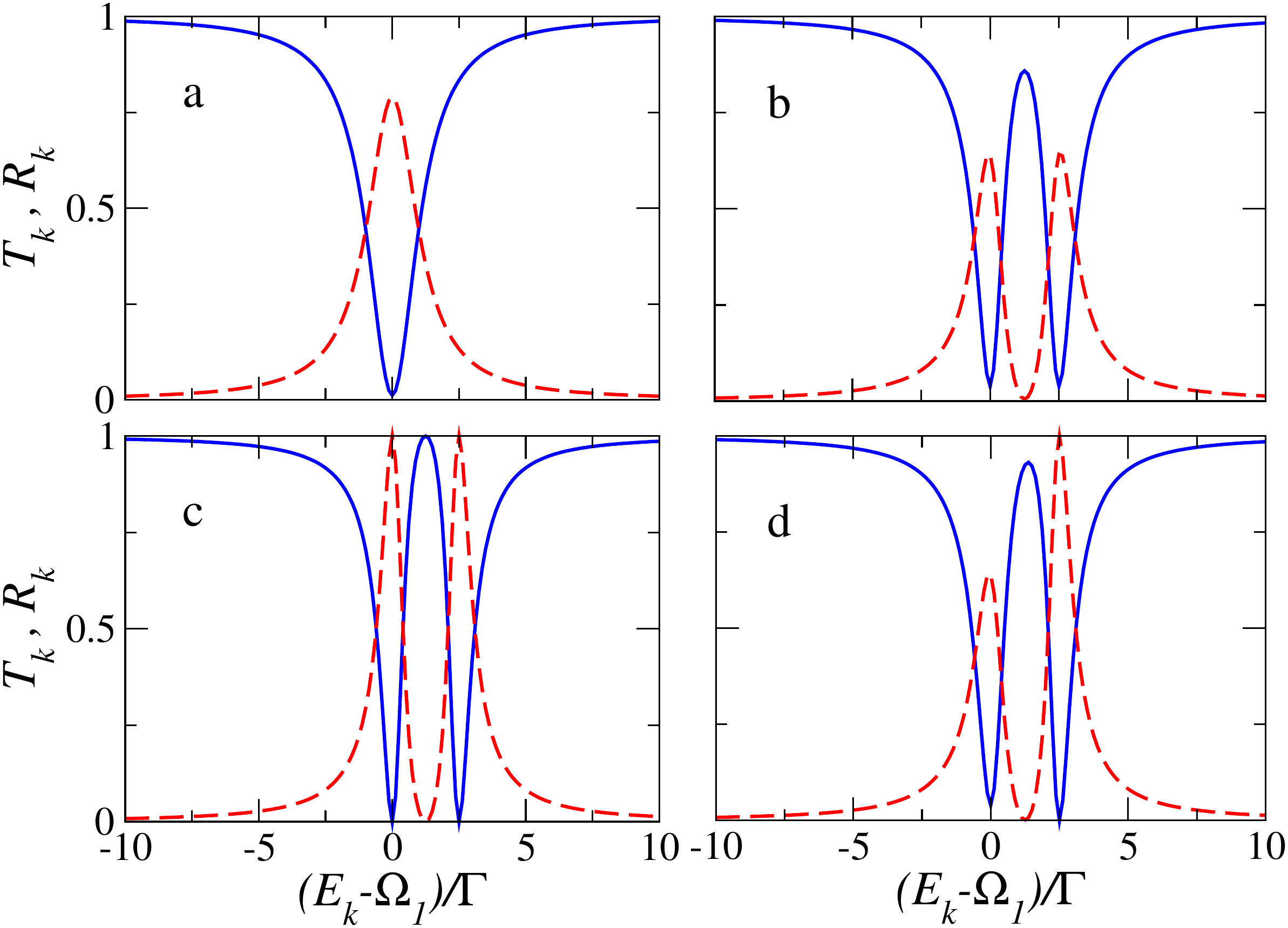}
\caption{Single-photon transmission $T_k$ (solid blue line) and reflection $R_k$  (dashed red line) coefficients for two 2LAs strongly coupled to the guided photon modes. In all panels $\Omega_2-\Omega_1=2.5\Gamma$ except $\Omega_1=\Omega_2$ in panel (a). The loss terms $\g_1=\g_2=\Gamma/4$ in panels (a, b), $\g_1=\g_2=0$ in panel (c), and $\g_1=\Gamma/4,~\g_2=0$ in panel (d).}
\label{single}
\end{figure}

 We first calculate single and two-photon scattering states for two different 2LAs. Our method here can readily be generalized for many different 2LAs. A single-photon scattering state $|k\ra$ of an incident photon with energy $E_k=k$ is given by 
\bea
|k\ra&=&\int dx \big\{ A_1(g_k(x)a^{\dagger}_{e}(x)+\delta(x)e_{k,1}\sigma_{1+}+\delta(x)e_{k,2}\sigma_{2+})\nn\\&+&B_1h_k(x)a^{\dagger}_o(x)\big\}|\varnothing \ra~,\label{sphs}
\eea
where $g_k(x)~(h_k(x))$ is an amplitude of a single photon in the even (odd) mode while $e_{k,l}$ is an amplitude of the excited $l$th atom with $l=1,2$. Here $A_1$ and $B_1$ satisfy initial conditions of the incident photon, i.e., the photon is incoming in the left- or the right-moving channel. $|\varnothing \ra$ represents vacuum state with the atoms in the ground state and zero photon in the left and right modes. We derive the amplitudes in Eq.\ref{sphs} using the single-photon Schr{\"o}dinger equation (see Appendix \ref{2atoms}). The amplitudes are $e_{k,1}=V(E_k-\tilde{\Omega}_2)/(\sqrt{2\pi} \Xi),~e_{k,2}=V(E_k-\tilde{\Omega}_1)/(\sqrt{2\pi} \Xi)$, $g_k(x)=(\theta(-x)+t_2(k)\theta(x))e^{ikx}/\sqrt{2\pi}$ and $h_k(x)=e^{ikx}/\sqrt{2\pi}$ where $\Xi=(E_k-\tilde{\Omega}_1+i\Gamma/2)(E_k-\tilde{\Omega}_2+i\Gamma/2)+\Gamma^2/4$, $\Gamma=V^2$ and 
\bea
t_2(k)=[(E_k-\tilde{\Omega}_1-i\f{\Gamma}{2})(E_k-\tilde{\Omega}_2-i\f{\Gamma}{2})+\f{\Gamma^2}{4}]/\Xi.\nn \\
\eea
The single-photon transmission coefficient $T_k=|t_2(k)+1|^2/4$, and reflection coefficient $R_k=|t_2(k)-1|^2/4$. We plot $T_k$ and $R_k$ in Fig.\ref{single}.  When two atoms are identical, $\Omega_1=\Omega_2=\Omega$ and $\g_1=\g_2=\g$, the excitation amplitude of the atoms, $e_{k,1}=e_{k,2}\equiv e_2(k)=\f{1}{\sqrt{2\pi}}\f{V}{E_k-\Omega+i\g/2+i\Gamma}$ and $t_2(k)\equiv \tau_2(k)=\f{E_k-\Omega+i\g/2-i\Gamma}{E_k-\Omega+i\g/2+i\Gamma}$. It has been shown earlier that a single 2LA in a 1D waveguide behaves as a perfect mirror, and the incident photon is completely reflected at single photon resonance, $E_k=\Omega$ and $\g=0$ \cite{Shen05}. It occurs because the spontaneous emission which is also part of the guided modes directly gives rise to the reflection. We find that the behaviour of the reflection and transmission coefficients of single photon for two identical 2LAs is similar to the single 2LA, i.e., $T_k=0$ and $R_k=1$ at single-photon resonance. $T_k$ is unity and $R_k$ is zero for two different 2LAs at $E_k=(\Omega_1+\Omega_2)/2$ and $\g_1=\g_2=0$. Incident photon at this energy rarely interacts with the atoms and thus transmits fully through the atoms. $R_k$ is unity and $T_k$ is zero for two different 2LAs when $E_k=\Omega_1$, $\g_1=0$ and/or $E_k=\Omega_2$, $\g_2=0$. This behaviour is again similar to the perfect mirror nature of a single atom as the incident photon does not interact with the other atom detuned in energy. The plots in Fig.\ref{single} show that a finite loss affects the single-photon reflection coefficient relatively more compared to the transmission coefficient. The reflection of photons in the side-coupled atom-photon system is solely due to emission of photons from the excited state of the atoms to the guided modes. Thus the loss of photons from the excited atomic state to non-guided modes reduces the reflection coefficient much more than the transmission coefficient. The transmission of photons becomes one when the incident photons do not interact with the atoms.  

The general two-photon scattering state of two incident photons with energy $E_{\bf k}=E_{k_1}+E_{k_2}=k_1+k_2$ has the form 
\bea
&&|k_1,k_2\ra=\int dx_1dx_2\Big[A_2\big\{g(x_1,x_2)\f{1}{\sqrt{2}}a^{\dg}_e(x_1)a^{\dg}_e(x_2)\nn\\&&+e_1(x_1)\delta(x_2)a^{\dg}_e(x_1)\sigma_{1+}+e_2(x_1)\delta(x_2)a^{\dg}_e(x_1)\sigma_{2+}\nn\\&&+e_3\delta(x_1)\delta(x_2)\sigma_{1+}\sigma_{2+}\big\}+B_2\big\{j(x_1;x_2)a^{\dg}_e(x_1)a^{\dg}_o(x_2)\nn \\&&+f_1(x_2)\delta(x_1)a^{\dg}_o(x_2)\sigma_{1+}+f_2(x_2)\delta(x_1)a^{\dg}_o(x_2)\sigma_{2+}\big\}\nn\\&&+C_2~h(x_1,x_2)\f{1}{\sqrt{2}}a^{\dg}_o(x_1)a^{\dg}_o(x_2)\Big]|\varnothing\ra~,
\label{wavefn}
\eea
%where the physical meaning of various amplitudes is obvious from the definition. 
where $e_1(x)~(f_1(x))$ and $e_2(x)~(f_2(x))$ are the amplitudes of the excited left and right atom respectively when there is a photon in the even (odd) mode. Here, $g(x_1,x_2),~j(x_1;x_2),$ and $h(x_1,x_2)$ are the amplitudes of two photons in the even modes, one in the even plus another in the odd mode, and two photons in the odd modes respectively. The amplitude of two excited atoms is given by $e_{3}$. The coefficients, $A_2,B_2$ and $C_2$ determine initial conditions of incident photons \cite{Roy10, Roy11}. We find these amplitudes from the two-photon Sch{\"o}dinger equation with a choice of incoming two-photon state (check Appendix \ref{2atoms}). Let us call $\tilde{\Omega}_1-\tilde{\Omega}_2=\Delta$, $\sqrt{\Delta^2-\Gamma^2}=\beta$ and $\lambda_{\pm}=E_{\bf k}-(\tilde{\Omega}_1+\tilde{\Omega}_2)/2+i\Gamma/2\pm\beta/2$. The amplitudes of two-photon wavefunction are
\bea
&&g(x_1,x_2)=\f{1}{\sqrt{2}}(g_{k_1}(x_1)g_{k_2}(x_2)+g_{k_2}(x_1)g_{k_1}(x_2))\nn\\&&+\Big[\theta(x_1-x_2)\theta(x_2)\Big(\f{\Delta+\beta-i\Gamma}{\sqrt{2}V}c_1e^{i\lambda_-x_1}e^{i(E_{\bf k}-\lambda_-)x_2}\nn\\&&+\f{\Delta-\beta-i\Gamma}{\sqrt{2}V}c_2e^{i\lambda_+x_1}e^{i(E_{\bf k}-\lambda_+)x_2}\Big)+(x_1 \leftrightarrow x_2)\Big],\label{g2}\\
&&j(x_1;x_2)=(g_{k_1}(x_1)h_{k_2}(x_2)+g_{k_2}(x_1)h_{k_1}(x_2)),\nn\\
&&h(x_1,x_2)=\f{1}{\sqrt{2}}(h_{k_1}(x_1)h_{k_2}(x_2)+h_{k_2}(x_1)h_{k_1}(x_2)).\nn
\eea
The part of the wavefunction in Eq.\ref{g2} involving $c_1,c_2$ (Appendix \ref{2atoms}) is generated because of inelastic photon scattering by the atoms, and is a signature of the background fluorescence. This part is also responsible for the photon-photon correlations created by resonant interactions of the light beam with the atoms. By integrating out the field operators of the 2LAs from Eq.\ref{wavefn} we  write down an asymptotic outgoing scattering state using the original right and left moving free photons as a combination of two transmitted $t_2(x_1,x_2)$, two reflected $r_2(x_1,x_2)$ and one transmitted plus one reflected photon $rt(x_1,x_2)$. For two incident photons in the right-moving channel, the asymptotic state is \cite{Roy11}
\bea
&&\int dx_1dx_2\Big[t_2(x_1,x_2)\f{1}{\sqrt{2}}a^{\dg}_R(x_1) a^{\dg}_R(x_2)+r_2(x_1,x_2)\nn \\&&\f{1}{\sqrt{2}}a^{\dg}_L(x_1) a^{\dg}_L(x_2)+rt(x_1,x_2)a^{\dg}_R(x_1) a^{\dg}_L(x_2)\Big]|\varnothing \ra~.\label{asymstate} 
\eea
Later we discuss nature of the various parts of the two-photon wavefunction in Eq.\ref{asymstate} of two different 2LAs in comparison  with that of a $V$-type 3LA. 
\begin{figure}
\includegraphics[width=8.5cm]{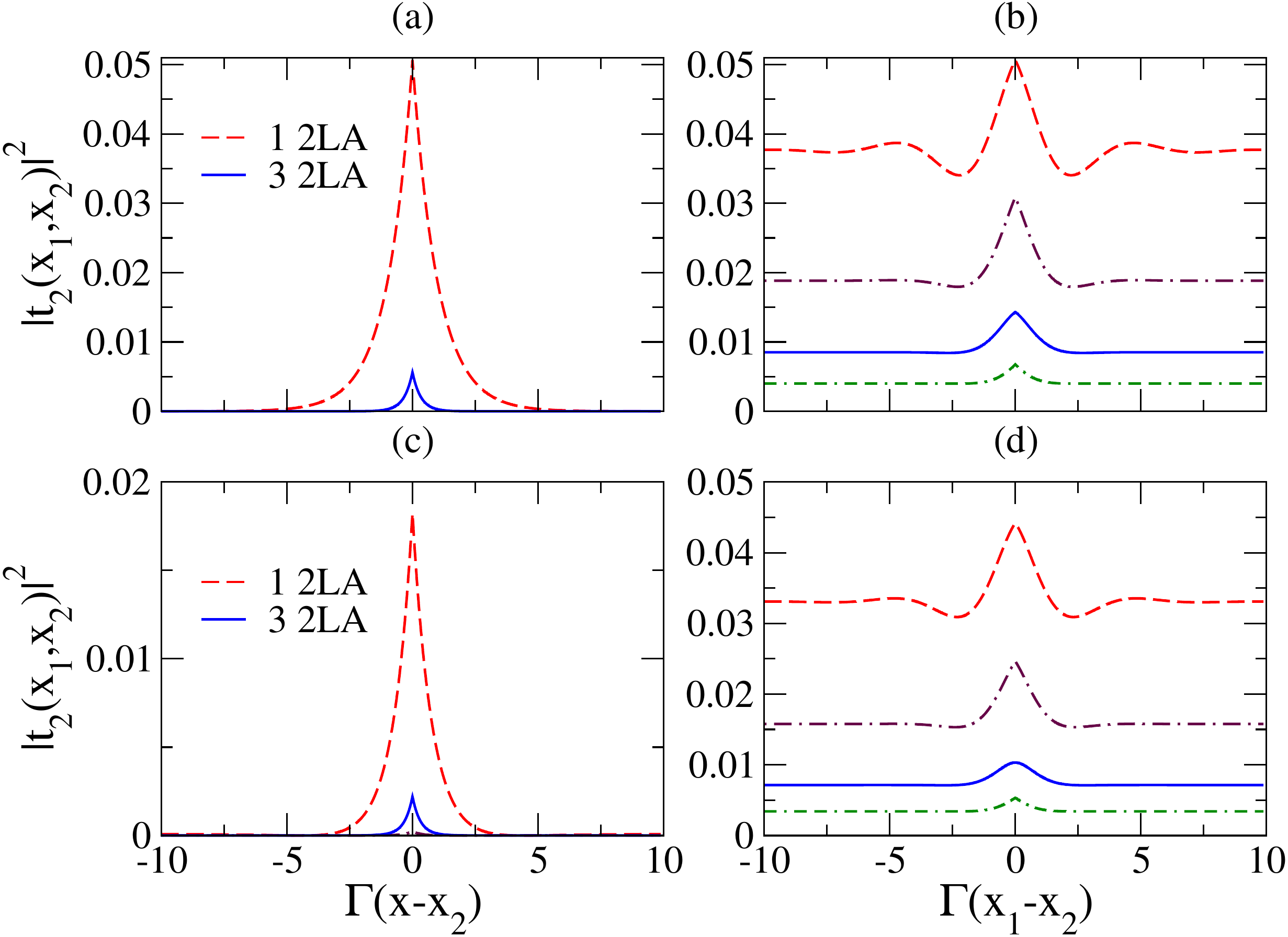}
\caption{Correlation of two transmitted photons $|t_2(x_1,x_2)|^2$ for multiple identical 2LAs with scaled distance separation $\Gamma(x_1-x_2)$ between photons. The left figures (a,c) are for two incident photons at single-photon resonance, $E_{k_1}=E_{k_2}=\Omega$. $|t_2(x_1,x_2)|^2$ is zero for even number of 2LAs at $E_{k_1}=E_{k_2}=\Omega$. The right figures (b,d) are at a detuned incident energy, $E_{k_1}=E_{k_2}$ and $E_{k_1}-\Omega=1.25\Gamma$. The number of 2LAs in the right figures (b, d) is 1 to 4 for the curves from the top to bottom. The loss $\g=0$ in (a,b) and $\g=\Gamma/4$ in (c,d).}
\label{multiple}
\end{figure}

\section{Multiple identical two-level atoms} 
\label{multi2la}
For two identical 2LAs, the expression in Eq.\ref{g2} of the two-photon wavefunction in the even modes reduces to 
\bea
&&g_2(x_1,x_2)=\f{1}{\sqrt{2}}(g_{2,k_1}(x_1)g_{2,k_2}(x_2)+g_{2,k_2}(x_1)g_{2,k_1}(x_2))\nn\\
&&-\f{iV}{\sqrt{2}}\mathcal{C}_2\big[e^{iE_{\bf k}x_2-i(\Omega-i\gamma/2-i\Gamma)(x_2-x_1)}\theta(x_2-x_1)\theta(x_1)+\nn \\
&&(x_1 \leftrightarrow x_2)\big],~{\rm where}~~g_{2,k}(x)=\f{e^{ikx}}{\sqrt{2\pi}}(\theta(-x)+\tau_2(k)\theta(x)),\nn\\
&&\mathcal{C}_2=8iVe_2(k_1)e_2(k_2)-\f{2\sqrt{2}i\Gamma(e_2(k_1)+e_2(k_2))}{\sqrt{\pi}(E_{\bf k}-2\Omega+i\g+i\Gamma)}.\label{g22}
\eea
Here $\mathcal{C}_2$ is a measure of the strength of photon-photon interactions. The form of the two-photon wavefunction in the even modes for multiple identical 2LAs is almost similar to the one in Eq.\ref{g22} except the spread of two-photon bound state (part of the wavefunction with $\mathcal{C}_2$) over $x_1-x_2$ reduces with increasing number of atoms. It signals the fall of two photon correlations with an increasing number of identical atoms. For example, the two-photon wavefunction in the even modes for three identical 2LAs is (Appendix \ref{3atoms})
\bea
&&g_3(x_1,x_2)=\f{1}{\sqrt{2}}(g_{3,k_1}(x_1)g_{3,k_2}(x_2)+g_{3,k_2}(x_1)g_{3,k_1}(x_2))\nn\\
&&-\f{iV}{\sqrt{2}}\mathcal{C}_3\big[e^{iE_{\bf k}x_2-i(\Omega-i\gamma/2-1.5i\Gamma)(x_2-x_1)}\theta(x_2-x_1)\theta(x_1)+\nn \\
&&(x_1 \leftrightarrow x_2)\big],~{\rm where}~~g_{3,k}(x)=\f{e^{ikx}}{\sqrt{2\pi}}(\theta(-x)+\tau_3(k)\theta(x)),\nn\\
&&\mathcal{C}_3=18iVe_3(k_1)e_3(k_2)-\f{6\sqrt{2}i\Gamma}{\sqrt{\pi}}\f{e_3(k_1)+e_3(k_2)}{E_{\bf k}-2\tilde{\Omega}+2i\Gamma},\label{g23}
\eea
and $e_3(k)=V/(\sqrt{2\pi}(E_k-\tilde{\Omega}+1.5i\Gamma)),~\tau_3(k)=(E_k-\tilde{\Omega}-1.5i\Gamma)/(E_k-\tilde{\Omega}+1.5i\Gamma)$. We find for four identical  2LAs (Appendix \ref{4atoms}), 
\bea
\mathcal{C}_4=32iVe_4(k_1)e_4(k_2)-\f{12\sqrt{2}i\Gamma}{\sqrt{\pi}}\f{e_4(k_1)+e_4(k_2)}{E_{\bf k}-2\tilde{\Omega}+3i\Gamma},\label{42LA}
\eea
where $e_4(k)=V/(\sqrt{2\pi}(E_k-\tilde{\Omega}+2i\Gamma))$ and $\tau_4(k)=(E_k-\tilde{\Omega}-2i\Gamma)/(E_k-\tilde{\Omega}+2i\Gamma)$. At single photon resonance, i.e., $E_k=\Omega$ and $\g=0$, we find $\mathcal{C}_l=0$ for an even number of identical 2LAs, and the magnitude of $\mathcal{C}_l$ falls with an increasing odd number of identical 2LAs (see Fig.\ref{multiple}(a)). Two degenerate incident photons at single-photon resonance are reflected from an even number of identical 2LAs by a purely elastic scattering process, thus no inelastic two-photon bound-state is formed in this case. Of course, $\mathcal{C}_l$ is not exactly zero even for an even number of atoms in a more physical situations with a finite value of loss (i.e., $\g \ne 0$) and a slightly detuned incident energy from the single photon resonance. We plot correlations of two transmitted photons in multiple identical 2LAs for incident photons at single-photon resonance in Figs.\ref{multiple}(a, c) and detuned from single-photon resonance in Figs.\ref{multiple}(b, d). The two-photon correlations fall monotonically with an increasing number of 2LAs for two detuned incident photons. In Figs.\ref{multiple}(c, d) we show that the nonmonotonic and monotonic behaviour of two-photon correlations respectively at single-photon resonance and away from single-photon resonance survive for a finite value of spontaneous emissions (i.e., $\g \ne 0$) to the non-guided modes.

\section{$V$-type three level atom (3LA)}
\label{V3LA}
 Next we consider a $V$-type 3LA as shown in Fig.\ref{3LA}(b). The Hamiltonian of a single 3LA in a 1D waveguide with right and left-moving photon modes is exactly identical to Eq.\ref{Ham} with $N=2$. The single-photon transport in a $V-$type 3LA is the same to that in the two 2LAs with similar transition energies. This is because single photon can excite only one atomic transition of two 2LAs or a 3LA at a time. Two photons can simultaneously excite two atomic transitions of two independent 2LAs, but not two transitions of a $V$-type 3LA where only one transition can be excited at a time. This creates a difference in two-photon transport in two 2LAs and in a $V$-type 3LA with similar energy configuration.

\begin{figure}
\includegraphics[width=8.5cm]{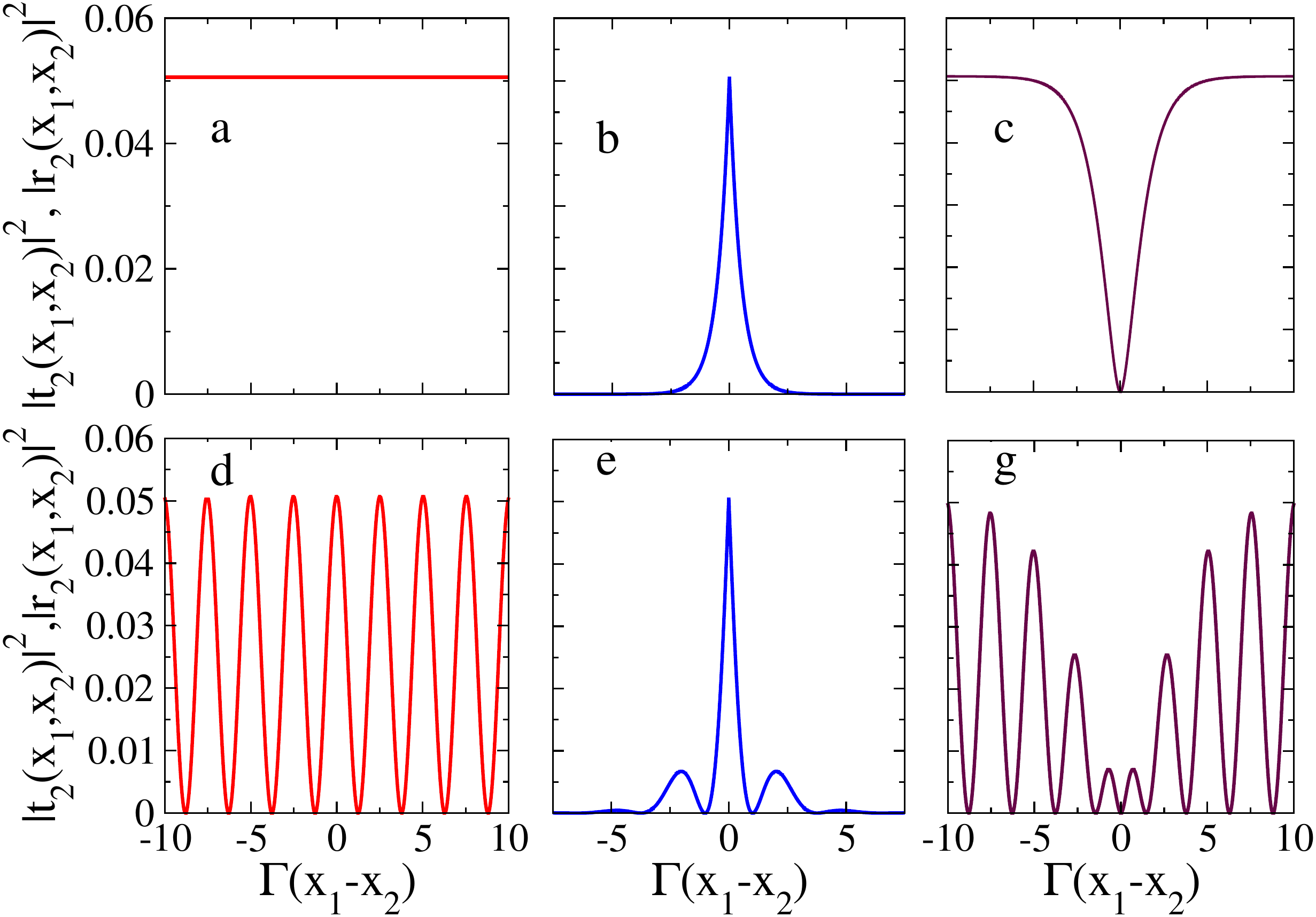}
\caption{Correlations of two transmitted $|t_2(x_1,x_2)|^2$ (middle column) and two reflected $|r_2(x_1,x_2)|^2$ (first and last columns) photons with scaled distance separation $\Gamma(x_1-x_2)$ between photons. The first column is for two 2LAs, and the middle and last columns are for a $V$-type 3LA. The parameters are, $\Omega_1 \to \Omega_2=E_{k_1}=E_{k_2}$ (first row) and $\Omega_1=E_{k_1},~\Omega_2=E_{k_2},~\Omega_2-\Omega_1=2.5\Gamma$ (second row). In all plots, $\g_1=\g_2=0$.}
\label{two}
\end{figure}

The two-photon transport in a $V$-type 3LA in a 1D waveguide can be found easily from the previous calculation for two different 2LAs by setting $e_3=0$ in Eq.\ref{wavefn}, because two excited levels of a 3LA can not be excited simultaneously. We compare nature of two transmitted and two reflected photons from two 2LAs and a 3LA in a 1D waveguide for different parameters. For simplicity we first consider $\g_1=\g_2=0$ in the following comparison, though a loss ($\g_l \ne 0$) develops little different features as shown later. (i) First, we consider the case when two transitions of two 2LAs or a 3LA are nearly identical, $\Omega_1 \to \Omega_2=\Omega$. It can occur both in real and artificial atoms. For example, energy level structure of a neutral quantum dot (such as, an InAs quantum dot) under a magnetic field includes a ground state and two bright exciton states which represent two anti-aligned spin configurations of the electron and hole \cite{Kim13}. The applied magnetic field controls the energy difference $(\Omega_2-\Omega_1)$ between the bright exciton states, and the energy difference at low magnetic field can be much smaller than the broadening $\Gamma$ due to coupling with the guided photon modes, i.e., $(\Omega_2-\Omega_1)<<\Gamma$. Therefore the energy of the excited states are almost degenerate at low magnetic field. Similar degeneracy in the excited states can also arise in atomic systems when the hyperfine splitting is consumed by the widening of spectral line-widths by different mechanisms including the Doppler broadening. When $\Omega_1 \to \Omega_2=\Omega$, two degenerate incident photons at $E_{k_1}=E_{k_2}=\Omega$ are perfectly reflected from two identical 2LAs, and the correlations of two reflected photons $|r_2(x_1,x_2)|^2$ is constant (which is equal to $1/2\pi^2$) for any distance separation $x=x_1-x_2$ between them (see Fig.\ref{two}(a)). The corresponding wavefunction of two transmitted photons $|t_2(x_1,x_2)|^2$ is zero at all $x$. In contrast, two reflected photons at these parameters for a $V$-type 3LA is anti-bunched, (check Fig.\ref{two}(c)) while two transmitted photons are bunched together, Fig.\ref{two}(b). The anti-bunching of two reflected photons in a side-coupled $V$-type 3LA occurs because two photons can not be emitted simultaneously by the 3LA, and the reflected photons are solely due to emission from the 3LA. (ii) Two degenerate incident photons each with energy equal to $(\Omega_1+\Omega_2)/2$ are fully transmitted through two different 2LAs, and and the correlations of two transmitted photons is constant (equal to $1/2\pi^2$) at any $x$. The corresponding wavefunction of two reflected photons is zero at all $x$. The nature of two transmitted and two reflected photons for a 3LA at these parameter sets is identical to the two different 2LAs. (iii) The correlation of two transmitted photons for two different 2LAs in a 1D waveguide is zero at all $x$ when the incident photon energy $E_{k_1}=\Omega_1$ and $E_{k_2}=\Omega_2$ with $\Omega_1 \ne \Omega_2$. The correlation of two reflected photons for two different 2LAs at these parameters shows a cosine oscillation with $x$ (see Fig.\ref{two}(d)). The nature of two transmitted and two reflected photons for a 3LA at these parameter sets is different from the two different 2LAs. Two transmitted photons from a 3LA are bunched (see Fig.\ref{two}(e)) while two reflected photons are anti-bunched in Fig.\ref{two}(f). The correlation of two reflected photons from a 3LA at these parameters also oscillates with $x$ (see Fig.\ref{two}(f)). 

\begin{figure}
\includegraphics[width=8.5cm]{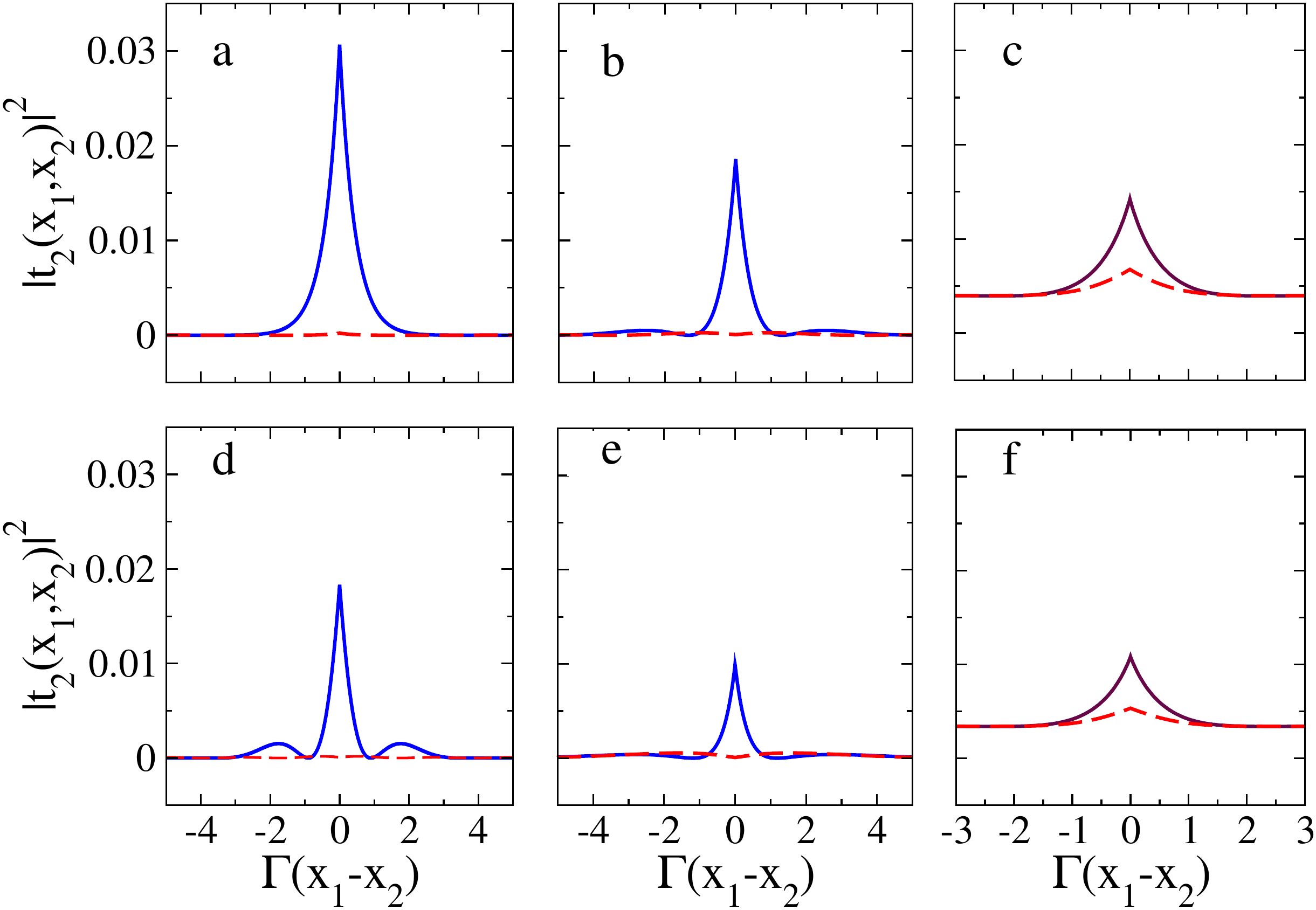}
\caption{Correlations of two transmitted photons $|t_2(x_1,x_2)|^2$  with scaled distance separation $\Gamma(x_1-x_2)$ between photons. The red dashed lines are for two (b,e) or four 2LAs (c,f), and the solid lines are for one (a,b,d,e) or two (c,f) $V$-type 3LAs. The parameters are, $\Omega_1 \to \Omega_2=E_{k_1}=E_{k_2}$in panel (a), $\Omega_1=E_{k_1},~\Omega_2=E_{k_2},~\Omega_2-\Omega_1=1.25\Gamma$  in panel (b), $\Omega_1 \to \Omega_2=E_{k_1}=E_{k_2}$ in panels (c, f), $\Omega_1=E_{k_1},~\Omega_2=E_{k_2},~\Omega_2-\Omega_1=2.5\Gamma$ in panel (d) and $\Omega_1=E_{k_1},~\Omega_2=E_{k_2},~\Omega_2-\Omega_1=0.5\Gamma$ in panel (e). In all panels $\g_1=\g_2=\Gamma/4$ except $\g_1=\g_2=0$ in panel (c) and $\g_1=\g_2=\Gamma/2$ in panel (e).}
\label{three}
\end{figure}

In all three above sets, the total energy of incident photons is $(\Omega_1+\Omega_2)$, when the fluorescence vanishes completely in two 2LAs, it has been discussed recently \cite{Rephaeli11, Muthukrishnan04}. However, a loss $\gamma_l \ne 0$ from the 2LAs can create non-zero fluorescence. The nature of two transmitted or two reflected photons in a system of two 2LAs and in a $V$-type 3LA are identical when the single photon transmission amplitude (which is same in both the systems) is exactly one. Then the amplitude $e_3$ of simultaneous excitation of both the 2LAs is zero, and the system of two 2LAs behaves like a $V$-type 3LA. Away from this special parameter set (which can be achieved easily with a slightly detuned incident photons or incorporating a loss $\gamma_l \ne 0$) the nature of two transmitted or two reflected photons in a system of two 2LAs and in a $V$-type 3LA are very different.  Therefore, two-photon scattering by a tightly focused weak light beam can be used to differentiate between two 2LAs and a $V$-type 3LA with similar transition energies. We show the correlations of two transmitted photons $|t_2(x_1,x_2)|^2$ from two 2LAs and a single $V$-type 3LA in Figs.\ref{three}(a,b,d,e) for finite values of the spontaneous emissions to non-guided modes, $\gamma_l \ne 0$. The differences in the two-photon correlations coming from two 2LAs and a 3LA remain pronounced even in the presence of loss from the excited states of the atoms. The  correlations of two transmitted photons in two 2LAs is almost zero at single-photon resonance for a finite loss.   

The single photon transmission in an ensemble of four 2LAs is similar to that in an ensemble of  two $V$-type 3LAs with similar level structures, for example, a 2LA ensemble where two 2LAs have transition energy $\Omega_1$ and other two have $\Omega_2$, while each 3LA in a 3LA ensemble has two transition energies $\Omega_1$ and $\Omega_2$. Now, two $V$-type 3LAs are saturated by two photons because two photons can simultaneously excite two transitions of two different 3LAs. Still, two-photon transport in four 2LAs is different from that in two 3LAs. This is because two photons can excite any two atoms of four 2LAs, thus there are total six possible amplitudes of exciting any two 2LAs, while two transitions from two different 3LAs can be excited by two photons, therefore there are total four options for two simultaneous atomic transitions in the 3LAs. It creates a difference between two-photon transport in four 2LAs and two-photon transport in two $V$-type 3LAs. Let us consider a system of four identical 2LAs for which we have already calculated the inelastic scattering contribution (see Eq.\ref{42LA}). The inelastic contribution in two-photon scattering from two identical $V$-type 3LAs with two almost similar transition energies $\Omega_1 \to \Omega_2=\Omega$ is given by 
\bea
\mathcal{C}_{2V}=32iVe_4(k_1)e_4(k_2)-\f{8\sqrt{2}i\Gamma}{\sqrt{\pi}}\f{e_4(k_1)+e_4(k_2)}{E_{\bf k}-2\tilde{\Omega}+2i\Gamma}\label{23LA}
\eea
where $e_4(k)$ and $\tau_4(k)$ are as before. Away from single-photon resonance, $\mathcal{C}_{2V} \ne \mathcal{C}_4$. The correlations of two transmitted photons in four identical 2LAs and two identical $V$-type 3LAs are shown in Figs.\ref{three}(c,f) in the absence and presence of loss from the excited states. The difference in two-photon scattering also persists between six 2LAs and three $V$-type 3LAs. However, with an increasing number of atoms this difference becomes weaker. Single photon transport in a driven $\Lambda$ type 3LA \cite{Witthaut10,Roy11} where one transition is coupled to photon modes and another transition is driven by a classical laser beam, is also similar to the two 2LAs in a 1D waveguide. We find that two-photon transport in a driven $\Lambda$ type 3LA matches with that in a $V$-type 3LA in a 1D waveguide.  

\section{Dipole-dipole interactions}
\label{DipoleI}
\begin{figure}
\includegraphics[width=8.5cm]{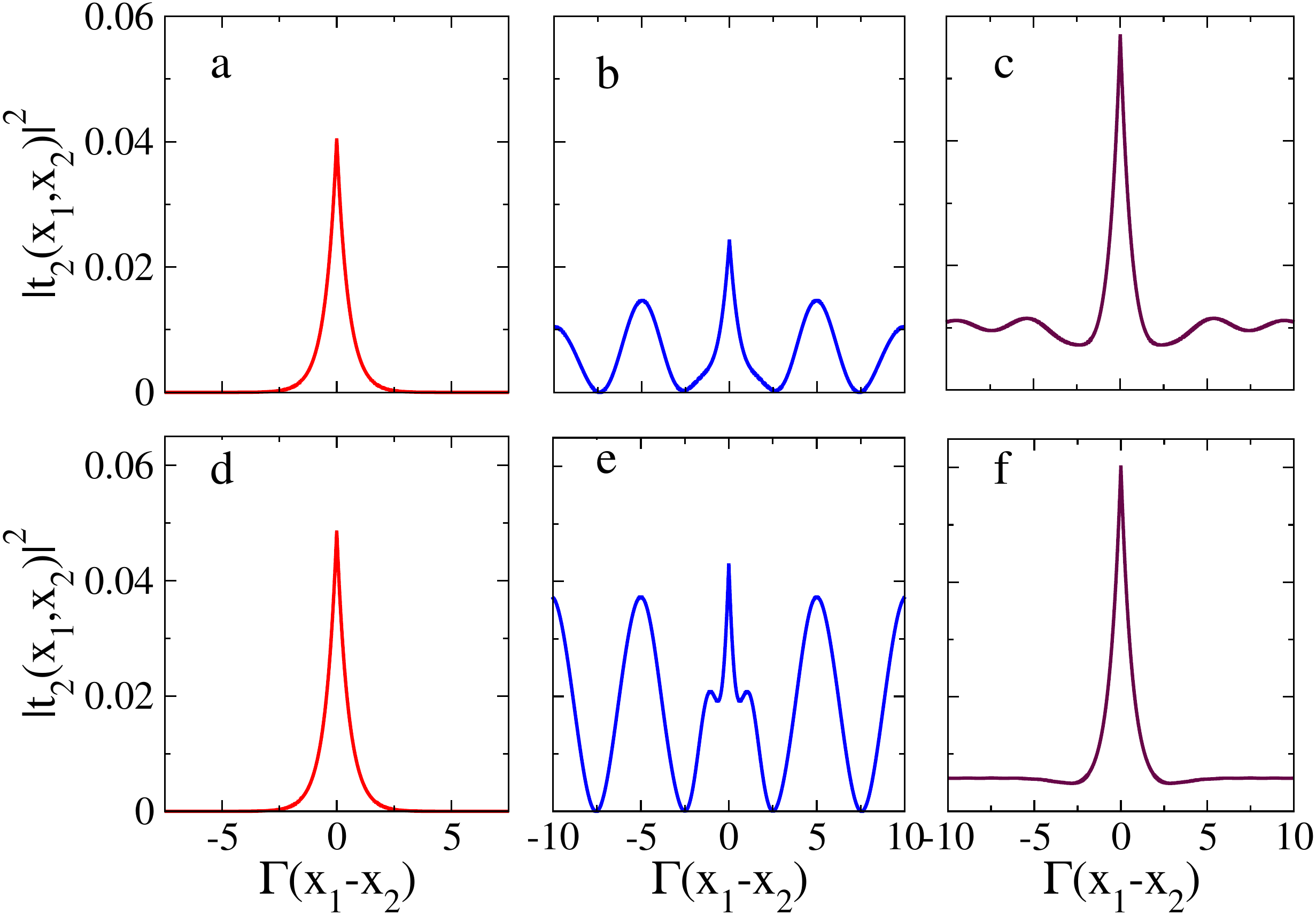}
\caption{Correlations of two transmitted photons $|t_2(x_1,x_2)|^2$  with scaled distance separation $\Gamma(x_1-x_2)$ between photons for two 2LAs at a finite dipole-dipole coupling $J$. The parameters are, $\Omega_1= \Omega_2$, $E_{k_1}=E_{k_2}=\Omega_1+J$ in panels (a,d), $\Omega_1=E_{k_1},~\Omega_2=E_{k_2},~\Omega_2-\Omega_1=1.25\Gamma$  in panels (b,e), and $\Omega_2-\Omega_1=1.25\Gamma,~E_{k_1}=E_{k_2}=\Omega_1+J$ in panels (c, f). The dipole-dipole coupling between atoms $J=\Gamma$ in the first row and $J=2.5\Gamma$ in the second row.}
\label{dipole}
\end{figure}

When we include a dipole-dipole interaction term $J\sum_{l,m=1}^{N}(\sigma_{l+}\sigma_{m-}+\sigma_{m-}\sigma_{l+})$ between 2LAs, the nature of two-photon correlations gets modified. For example, a single photon is fully reflected from two identical 2LAs when energy of incident photon is $E_k=\Omega+J$ and $\gamma=0$ where $\Omega_1=\Omega_2=\Omega$. However the correlation of two transmitted photons through two identical 2LAs is nonzero for a finite $J$ at single-photon resonance, $E_{k_1}=E_{k_2}=\Omega+J$ as shown in Fig.\ref{dipole}(a,d). The correlation of two transmitted photons reduces to zero as the dipole-dipole interaction vanishes (see Sec.\ref{multi2la}). We also find that the photon-photon correlation due to two correlated 2LAs can be stronger than the maximum photon-photon correlation generated by a single 2LA. It is shown in Fig.\ref{dipole}(c,f) where the height of peaks is larger than the height of the peaks in Fig.\ref{multiple}(a,b). Notice that we choose energy of the incident photons at single-photon resonance of one atom in Fig.\ref{multiple}(a,b). Correlation of two transmitted photons oscillates with the distance between atoms when energy of the both incident photons are away from single-photon resonance of any atom.

\section{Discussion and prospects}
\label{diss}
Resonance fluorescence of two two-level atoms in  running- and standing-wave laser fields has been discussed before for studying various interesting phenomena ranging from quantized cooling of identical atoms in the laser fields to inelastic coherent back-scattering spectrum of the laser light incident on cold atoms \cite{Guo95, Rudolph95, Lenz93, Shatokhin07}. Most of these studies are based on the master equations where the quantum regression theorem is used to derive resonance fluorescence spectrum emitted by these atoms. While an exact fully microscopic analysis is beyond the scope of these earlier studies, the effects arising from the distance separation between two two-level atoms are investigated carefully. The main difference between these earlier studies in 1D and the present study is the inclusion of spontaneous emission from atoms into the guided modes in our study. The two-photon scattering from multiple two-level atoms have been evaluated in some recent papers \cite{Yudson08, Pletyukhov12}. However none of these studies has investigated carefully a scaling of two-photon nonlinearity with an increasing number of two-level atoms at single-photon resonance and away from single-photon resonance. The differences in the resonance fluorescence of two-level atoms and three-level atoms in a one-dimensional tightly focused laser field have not been discussed earlier according to our best knowledge. The inelastic two-photon scattering of a tightly focused laser beam can be used to differentiate two  and three-level atoms with similar level structures which have similar line-shape in the single-photon scattering. Therefore, it can be viewed as some type of spectroscopy based on inelastic scattering such as Raman spectroscopy. The input two-photon Fock state can be generated using a weak coherent state input or using deterministic single-photon sources \cite{Kuhn02} and single-photon pulses \cite{Kolchin08}. The deterministic creation of pure two-photon Fock states has been demonstrated in cavity quantum electrodynamics \cite{Varcoe00, Bertet02} and solid-state systems \cite{Hofheinz08}. Many practical challenges relevant in experimental study of the present system have been discussed in Ref.\cite{Rephaeli11}. The theoretical method employed in this paper can be further extended to study scattering of three or more photons in this geometry. One would also be able to include the distance separations between atoms within this scattering theory approach.      

\section{Acknowledgments}
The support of the U.S. Department of Energy through LANL/LDRD Program for this work is gratefully acknowledged. 
\appendix
\section{Single and two-photon scattering by two different two-level atoms}
\label{2atoms}
In this appendix we provide a derivation of the single and two-photon scattering states of a tightly focused weak light beam. We start with the transformed Hamiltonian of the atom-photon system obtained in the main text after the even-odd transformation. The Hamiltonian reads, 
\bea
&&\mathcal{H}=\mathcal{H}_e+\mathcal{H}_o,~{\rm where}\label{Hamt}\\
&&\mathcal{H}_e+\mathcal{H}_o=-i v_g\int dx ~(a^{\dagger}_{e}(x)\partial_xa_{e}(x)+a^{\dagger}_{o}(x)\partial_xa_{o}(x))\nn\\&&+\sum_{l=1}^{2}\big[V(a_e^{\dg}(0)\sigma_{l-}+H.c.)+\tilde{\Omega}_la^{\dg}_{el}a_{el}\big],
\eea
with $\tilde{\Omega}_l=\Omega_l-i\gamma_l/2$. We set the group velocity $v_g=1$.  

{\it Single-photon scattering state:} A single-photon scattering state $|k\ra$ of an incident photon with energy $E_k=k$ is given in Eq.\ref{sphs}. 
%\bea
%|k\ra&=&\int dx \big\{ A_1(g_k(x)a^{\dagger}_{e}(x)+\delta(x)e_{k,1}\sigma_{1+}+\delta(x)e_{k,2}\sigma_{2+})\nn\\&+&B_1h_k(x)a^{\dagger}_o(x)\big\}|\varnothing \ra.\label{sphs}
%\eea
 We derive the amplitudes in Eq.\ref{sphs} using the single-photon Schr{\"o}dinger equation, $\mathcal{H}|k\ra=E_k|k\ra$. Thus we obtain the following linear equations for the amplitudes in $|k\ra$ from the single-photon Schr{\"o}dinger equation,
\bea
-i\partial_xg_k(x)-E_kg_k(x)+V\delta(x)(e_{k,1}+e_{k,2})&=&0,\label{tsse1}\\
(\Omega_1-i\f{\g_1}{2}-E_k)e_{k,1}+Vg_k(x)\delta(x)&=&0,\label{tsse2}\\
(\Omega_2-i\f{\g_2}{2}-E_k)e_{k,2}+Vg_k(x)\delta(x)&=&0.\label{tsse3}
\eea
We find a continuity relation across $x=0$, $g_k(0+)=g_k(0-)-iV(e_{k,1}+e_{k,2})$ from Eq.\ref{tsse1}. We use the regularization, $g_k(0)=(g_k(0+)+g_k(0-))/2$, and the initial condition $g_k(x<0)=e^{ikx}/\sqrt{2\pi}$. Thus we find from Eqs.\ref{tsse2},\ref{tsse3}
\bea
(E_k-\Omega_1+\f{i}{2}(\g_1+V^2))e_{k,1}+\f{iV^2}{2}e_{k,2}&=&\f{V}{\sqrt{2\pi}},\nn\\
(E_k-\Omega_2+\f{i}{2}(\g_2+V^2))e_{k,2}+\f{iV^2}{2}e_{k,1}&=&\f{V}{\sqrt{2\pi}}.\nn
\eea
Solving the above two equations we find $e_{k,1}=V(E_k-\tilde{\Omega}_2)/(\sqrt{2\pi} \Xi),~e_{k,2}=V(E_k-\tilde{\Omega}_1)/(\sqrt{2\pi} \Xi)$ where $\Xi=(E_k-\tilde{\Omega}_1+i\Gamma/2)(E_k-\tilde{\Omega}_2+i\Gamma/2)+\Gamma^2/4$ and $\Gamma=V^2$. Using $e_{k,1},~e_{k,2}$ in the continuity relation of $g_k(x)$ across $x=0$, we find 
 $g_k(x)=(\theta(-x)+t_2(k)\theta(x))e^{ikx}/\sqrt{2\pi}$ with
\bea
t_2(k)=\f{(E_k-\tilde{\Omega}_1-i\Gamma/2)(E_k-\tilde{\Omega}_2-i\Gamma/2)+\Gamma^2/4}{(E_k-\tilde{\Omega}_1+i\Gamma/2)(E_k-\tilde{\Omega}_2+i\Gamma/2)+\Gamma^2/4}.\nn
\eea
As a photon in the odd mode does not interact with the atoms, we find $h_k(x)=e^{ikx}/\sqrt{2\pi}$.

{\it Two-photon scattering state:}
  A general two-photon scattering state of two incident photons is written in Eq.\ref{wavefn}. %with energy, $E_{\bf k}=E_{k_1}+E_{k_2}=k_1+k_2$ has the form 
%\begin{widetext}
%\bea
%&&|k_1,k_2\ra=\int dx_1dx_2\Big[A_2\big\{g(x_1,x_2)\f{1}{\sqrt{2}}a^{\dg}_e(x_1)a^{\dg}_e(x_2)+e_1(x_1)\delta(x_2)a^{\dg}_e(x_1)\sigma_{1+}+e_2(x_1)\delta(x_2)a^{\dg}_e(x_1)\sigma_{2+}+e_3\delta(x_1)\delta(x_2)\sigma_{1+}\sigma_{2+}\big\}\nn\\&&+B_2\big\{j(x_1;x_2)a^{\dg}_e(x_1)a^{\dg}_o(x_2)+f_1(x_2)\delta(x_1)a^{\dg}_o(x_2)\sigma_{1+}+f_2(x_2)\delta(x_1)a^{\dg}_o(x_2)\sigma_{2+}\big\}+C_2~h(x_1,x_2)\f{1}{\sqrt{2}}a^{\dg}_o(x_1)a^{\dg}_o(x_2)\Big]|\varnothing\ra,
%\label{wavefun}
%\eea
%\end{widetext}
%where $e_1(x)~(f_1(x))$ and $e_2(x)~(f_2(x))$ are the amplitudes of the excited left and right atom respectively when there is a photon in the even (odd) mode. Here, $g(x_1,x_2),~j(x_1;x_2),$ and $h(x_1,x_2)$ are the amplitudes of two photons in the even modes, one in the even plus another in the odd mode, and two photons in the odd modes respectively. The amplitude of two excited atoms is given by $e_{3}$. 
From the two-photon Sch{\"o}dinger equation, $\mathcal{H}|k_1,k_2\ra=E_{\bf k}|k_1,k_2\ra$, we find the following linear differential equations for the amplitudes in Eq.\ref{wavefn},
\begin{widetext}
\bea
\Big(-i\partial_{x_1}-i\partial_{x_2}-E_{\bf k}\Big)g(x_1,x_2)+\f{V}{\sqrt{2}}\sum_{l=1,2}[e_l(x_1)\delta(x_2)+\delta(x_1)e_l(x_2)]&=&0, \label{tse1} \\
\Big(-i\partial_x-E_{\bf k}+\Omega_1-i\f{\g_1}{2}\Big)e_1(x)+\f{V}{\sqrt{2}}[g(0,x)+g(x,0)]+Ve_3\delta(x)&=&0, \label{tse2}  \\
\Big(-i\partial_x-E_{\bf k}+\Omega_2-i\f{\g_2}{2}\Big)e_2(x)+\f{V}{\sqrt{2}}[g(0,x)+g(x,0)]+Ve_3\delta(x)&=&0, \label{tse3}  \\
(\Omega_1-i\f{\g_1}{2}+\Omega_2-i\f{\g_2}{2}-E_{\bf k})e_3+V(e_1(0)+e_2(0))&=&0, \label{tse4} \\
\Big(-i\partial_{x_1}-i\partial_{x_2}-E_{\bf k}\Big)j(x_1;x_2)+V\delta(x_1)(f_1(x_2)+f_2(x_2))&=&0, \label{tse5}
\eea
\end{widetext}
\bea
\Big(-i\partial_x-E_{\bf k}+\Omega_1-i\f{\g_1}{2}\Big)f_1(x)+Vj(0;x)&=&0, \label{tse6} \\
\Big(-i\partial_x-E_{\bf k}+\Omega_2-i\f{\g_2}{2}\Big)f_2(x)+Vj(0;x)&=&0, \label{tse7} \\
\Big(-i\partial_{x_1}-i\partial_{x_2}-E_{\bf k}\Big)h(x_1,x_2)&=&0. \label{tSe8}
\eea
The Eqs.\ref{tse2},\ref{tse3} are coupled to each other by $g(x_1,x_2)$, and  we write Eqs.\ref{tse2},\ref{tse3} using the matrix notation,
\bea
\partial_x\vec{e}(x)&=&i\overleftrightarrow{\bf A}\vec{e}(x)-\sqrt{2}iVg(x,0-)\vec{\bf 1}-iVe_3\delta(x)\vec{\bf 1},~{\rm where}\nn\\\overleftrightarrow{\bf A}&=&\left( \begin{array}{cc} E_{\bf k}-\tilde{\Omega}_1+i\f{\Gamma}{2}  & i\f{\Gamma}{2} \\ i\f{\Gamma}{2} & E_{\bf k}-\tilde{\Omega}_2+i\f{\Gamma}{2}  \end{array}\right),\label{mat2}
\eea
and $\vec{e}(x)=(e_1(x), e_2(x))^{\rm T}$, $\vec{\bf 1}=(1,1)^{\rm T}$ where $T$ stands for transpose. We can decouple the coupled linear-differential equations in Eq.\ref{mat2} by using the following transformation. We define here, $\tilde{\Omega}_1-\tilde{\Omega}_2=\Delta$, $\sqrt{\Delta^2-\Gamma^2}=\beta$. The eigenvalues of $\overleftrightarrow{\bf A}$ are $\lambda_{\pm}=E_{\bf k}-(\tilde{\Omega}_1+\tilde{\Omega}_2)/2+i\Gamma/2\pm\beta/2$. We form a $2\times2$ square matrix $\overleftrightarrow{\bf P}$ using the eigenvectors $|\lambda_{+}\ra,~|\lambda_{-}\ra$ of $\overleftrightarrow{\bf A}$, thus $\overleftrightarrow{\bf P}=(|\lambda_-\ra, |\lambda_+\ra)$, and we have
\bea
\overleftrightarrow{\bf P}^{-1}\overleftrightarrow{\bf A}\overleftrightarrow{\bf P}=\left( \begin{array}{cc} \lambda_-  & 0 \\ 0 &  \lambda_+ \end{array}\right).
\eea
Therefore, we write,
\bea
\partial_x\big(\overleftrightarrow{\bf P}^{-1}\vec{e}(x)\big)&=&i\big(\overleftrightarrow{\bf P}^{-1}\overleftrightarrow{\bf A}\overleftrightarrow{\bf P}\big)\big(\overleftrightarrow{\bf P}^{-1}\vec{e}(x)\big)\nn\\&-&\sqrt{2}iVg(x,0-)\overleftrightarrow{\bf P}^{-1}\vec{\bf 1}-iVe_3\delta(x)\overleftrightarrow{\bf P}^{-1}\vec{\bf 1},\nn\\\label{rotex}
\eea
which gives two decoupled linear-differential equations for the transformed amplitudes, $\vec{\tilde{e}}(x)=\overleftrightarrow{\bf P}^{-1}\vec{e}(x)$. A similar transformation decouples the Eqs.\ref{tse6},\ref{tse7}. Next we use the method of the Ref.\cite{Roy12} to calculate all the amplitudes in the Eq.
\ref{wavefn}.
We find 
\bea
e_1(x)&=&(\f{i(\Delta+\beta)}{\Gamma}c_1e^{i\lambda_-x}+\f{i(\Delta-\beta)}{\Gamma}c_2e^{i\lambda_+x})\theta(x)\nn\\&+&(e_{k_1,1}g_{k_2}(x)+e_{k_2,1}g_{k_1}(x)),\label{e1x}\\
e_2(x)&=&(c_1e^{i\lambda_-x}+c_2e^{i\lambda_+x})\theta(x)\nn\\&+&(e_{k_1,2}g_{k_2}(x)+e_{k_2,2}g_{k_1}(x)),\label{e2x}
\eea
and $f_l(x)=e_{k_1,l}h_{k_2}(x)+e_{k_2,l}h_{k_1}(x),$ with $l=1,2$, and $e_3=i(t_2(k_1)+t_2(k_2)-2)/(2\pi(E_{\bf k}-\tilde{\Omega}_1-\tilde{\Omega}_2+i\Gamma))$. Here 
\bea
c_1&=&\f{-i\Gamma-\Delta+\beta}{4\pi\beta}i(\Gamma\varepsilon_{k_1}\eta_{k_2}+\Gamma\varepsilon_{k_2}\eta_{k_1}-2\pi V e_3),\nn\\
c_2&=&\f{i\Gamma+\Delta+\beta}{4\pi\beta}i(\Gamma \varsigma_{k_1}\eta_{k_2}+ \Gamma\varsigma_{k_2}\eta_{k_1}-2\pi V e_3),
\eea
and $\eta_k=i(t_2(k)-1)/\Gamma$, $\varepsilon_k=V/(k-(\tilde{\Omega}_1+\tilde{\Omega}_2)/2+i\Gamma/2-\beta/2)$, $\varsigma_k=V/(k-(\tilde{\Omega}_1+\tilde{\Omega}_2)/2+i\Gamma/2+\beta/2)$. The amplitudes of two-photon wavefunction are given in Eq.\ref{g2} and next two equations.
%\bea
%g(x_1,x_2)&=&\f{1}{\sqrt{2}}(g_{k_1}(x_1)g_{k_2}(x_2)+g_{k_2}(x_1)g_{k_1}(x_2))+\Big[\Big(\f{\Delta+\beta-i\Gamma}{\sqrt{2}V}c_1e^{i\lambda_-x_1}e^{i(E_{\bf k}-\lambda_-)x_2}\nn\\&+&\f{\Delta-\beta-i\Gamma}{\sqrt{2}V}c_2e^{i\lambda_+x_1}e^{i(E_{\bf k}-\lambda_+)x_2}\Big)\theta(x_1-x_2)\theta(x_2)+(x_1 \leftrightarrow x_2)\Big],\label{g2}\\j(x_1;x_2)&=&(g_{k_1}(x_1)h_{k_2}(x_2)+g_{k_2}(x_1)h_{k_1}(x_2)),\\h(x_1,x_2)&=&\f{1}{\sqrt{2}}(h_{k_1}(x_1)h_{k_2}(x_2)+h_{k_2}(x_1)h_{k_1}(x_2)).
%\eea
%The part of the wavefunctions in Eqs.\ref{e1x},\ref{e2x},\ref{g2} involving $c_1,c_2$ is generated due to  inelastic photon scattering by the atoms, and is a signature of the background fluorescence. This part is also responsible for the photon-photon correlations created by resonant interactions of the light beam with the atoms.

\section{Two-photon scattering by three two-level atoms}
\label{3atoms}
The Hamiltonian of three two-level atoms coupled to a tightly focused weak light beam after the even-odd transformation for the photon modes reads, $\mathcal{H}=\mathcal{H}_e+\mathcal{H}_o$ where
\bea
\mathcal{H}_e+\mathcal{H}_o&=&-i v_g\int dx (a^{\dagger}_{e}(x)\partial_xa_{e}(x)+a^{\dagger}_{o}(x)\partial_xa_{o}(x))\nn\\&+&\sum_{l=1}^{3}\big[V(a_e^{\dg}(0)\sigma_{l-}+H.c.)+\tilde{\Omega}_la^{\dg}_{el}a_{el}\big].
\eea
We set the group velocity $v_g=1$ as before. A two-photon scattering state in the above system for two incident photons with energy, $E_{\bf k}=k_1+k_2$ is given by 
\begin{widetext}
\bea
&&|k_1,k_2\ra=\int\Big[A_2\big\{{\bf g}(x_1,x_2)\f{1}{\sqrt{2}}a^{\dg}_e(x_1)a^{\dg}_e(x_2)+({\bf e}_1(x_1)\sigma_{1+}+{\bf e}_2(x_1)\sigma_{2+}+{\bf e}_3(x_1)\sigma_{3+})\delta(x_2)a^{\dg}_e(x_1)\nn\\&&+\delta(x_1)\delta(x_2){\bf (e}_{12}\sigma_{1+}\sigma_{2+}+{\bf e}_{13}\sigma_{1+}\sigma_{3+}+{\bf e}_{23}\sigma_{2+}\sigma_{3+})\big\}+B_2\big\{{\bf j}(x_1;x_2)a^{\dg}_e(x_1)a^{\dg}_o(x_2)+({\bf f}_1(x_2)\sigma_{1+}\nn\\&&+{\bf f}_2(x_2)\sigma_{2+}+{\bf f}_3(x_2)\sigma_{3+})\delta(x_1)a^{\dg}_o(x_2)\big\}+C_2~{\bf h}(x_1,x_2)\f{1}{\sqrt{2}}a^{\dg}_o(x_1)a^{\dg}_o(x_2)\Big] dx_1dx_2|\varnothing \ra.
\label{3wavefn}
\eea
\end{widetext}
We use bold symbols for the amplitudes in this appendix to distinguish them from the previous appendix. We find a set of coupled linear-differential equations using the two-photon Sch{\"o}dinger equation,
\begin{widetext}
\bea
\Big(-i\partial_{x_1}-i\partial_{x_2}-E_{\bf k}\Big){\bf g}(x_1,x_2)+\f{V}{\sqrt{2}}\sum_{l=1}^{3}[{\bf e}_l(x_1)\delta(x_2)+\delta(x_1){\bf e}_l(x_2)]&=&0, \label{3tse1} \\
\Big(-i\partial_x-E_{\bf k}+\tilde{\Omega}_1\Big){\bf e}_1(x)+\f{V}{\sqrt{2}}[{\bf g}(0,x)+{\bf g}(x,0)]+V({\bf e}_{12}+{\bf e}_{13})\delta(x)&=&0, \label{3tse2}  \\
\Big(-i\partial_x-E_{\bf k}+\tilde{\Omega}_2\Big){\bf e}_2(x)+\f{V}{\sqrt{2}}[{\bf g}(0,x)+{\bf g}(x,0)]+V({\bf e}_{23}+{\bf e}_{12})\delta(x)&=&0, \label{3tse3}  \\
\Big(-i\partial_x-E_{\bf k}+\tilde{\Omega}_3\Big){\bf e}_3(x)+\f{V}{\sqrt{2}}[{\bf g}(0,x)+{\bf g}(x,0)]+V({\bf e}_{23}+{\bf e}_{13})\delta(x)&=&0, \label{3tse4}  \\
\Big(-i\partial_{x_1}-i\partial_{x_2}-E_{\bf k}\Big){\bf j}(x_1;x_2)+V\delta(x_1)({\bf f}_1(x_2)+{\bf f}_2(x_2)+{\bf f}_3(x_2))&=&0,\label{3tse8}
\eea
\end{widetext}
\bea
(\tilde{\Omega}_1+\tilde{\Omega}_2-E_{\bf k}){\bf e}_{12}+V({\bf e}_1(0)+{\bf e}_2(0))&=&0,\label{3tse5} \\
(\tilde{\Omega}_2+\tilde{\Omega}_3-E_{\bf k}){\bf e}_{23}+V({\bf e}_2(0)+{\bf e}_3(0))&=&0,\label{3tse6} \\
(\tilde{\Omega}_1+\tilde{\Omega}_3-E_{\bf k}){\bf e}_{13}+V({\bf e}_1(0)+{\bf e}_3(0))&=&0,\label{3tse7} \\
\Big(-i\partial_x-E_{\bf k}+\tilde{\Omega}_1\Big){\bf f}_1(x)+V{\bf j}(0;x)&=&0, \label{tse9} \\
\Big(-i\partial_x-E_{\bf k}+\tilde{\Omega}_2\Big){\bf f}_2(x)+V{\bf j}(0;x)&=&0, \label{tse10} \\
\Big(-i\partial_x-E_{\bf k}+\tilde{\Omega}_3\Big){\bf f}_3(x)+V{\bf j}(0;x)&=&0, \label{tse11} \\
\Big(-i\partial_{x_1}-i\partial_{x_2}-E_{\bf k}\Big){\bf h}(x_1,x_2)&=&0. \label{tSe12}
\eea
We can find solutions of the above coupled linear-differential equations following the method of the Appendix \ref{2atoms}. Here we are interested to study identical atoms, and simplify the problem by choosing, $\Omega_1=\Omega_2=\Omega_3=\Omega$ and $\g_{1}=\g_{2}=\g_{3}=\g$. We define for the identical atoms, ${\bf g}(x_1,x_2)\equiv g_3(x_1,x_2),~{\bf e}_1(x)+{\bf e}_2(x)+{\bf e}_3(x)=\tilde{\bf e}(x)$ and ${\bf e}_{12}+{\bf e}_{13}+{\bf e}_{23}={\boldsymbol \rho}$. Thus we find from Eqs.\ref{3tse1}-\ref{3tse7},
\bea
&&\Big(-i\partial_{x_1}-i\partial_{x_2}-E_{\bf k}\Big)g_3(x_1,x_2)+\f{V}{\sqrt{2}}\tilde{\bf e}(x)=0, \label{33tse1} \\
&&\Big(-i\partial_x-E_{\bf k}+\tilde{\Omega}\Big)\tilde{\bf e}(x)+\f{3V}{\sqrt{2}}[g_3(0,x)+g_3(x,0)]\nn\\
&&+2V{\boldsymbol \rho}\delta(x)=0, \label{33tse2} \\
&&(2\tilde{\Omega}-E_{\bf k}){\boldsymbol \rho}+2V\tilde{\bf e}(0)=0.\label{33tse3}
\eea
We get from the last equation ${\boldsymbol \rho}=2V\tilde{\bf e}(0)/(E_{\bf k}-2\tilde{\Omega})$. The discontinuity relations from the above relations are, $g_3(x,0+)=g_3(x,0-)-\f{iV}{\sqrt{2}}\tilde{\bf e}(x)$ and $\tilde{\bf e}(0+)=\tilde{\bf e}(0-)-2iV{\boldsymbol \rho}$. Now we  calculate $\tilde{\bf e}(x<0)$ from Eq.\ref{33tse2} using the initial condition of $g_3(x_1,x_2)$,
\bea
\tilde{\bf e}(x<0)&=&\f{3}{\sqrt{2\pi}}(e_3(k_2)e^{ik_1x}+e_3(k_1)e^{ik_2x}),~{\rm where}\nn\\e_3(k)&=&\f{V}{\sqrt{2\pi}(E_k-\tilde{\Omega}+1.5i\Gamma)}.
\eea
Next we derive $g_3(x_1,x_2)$ in the region $x_1<0,x_2>0$,
\bea
g_3(x_1,x_2)&=&\f{1}{2\pi\sqrt{2}}(e^{ik_1x_1+ik_2x_2}\tau_3(k_2)+e^{ik_2x_1+ik_1x_2}\tau_3(k_1)),\nn\\\tau_3(k)&=&1-3iVe_3(k)=\f{E_k-\tilde{\Omega}-1.5i\Gamma}{E_k-\tilde{\Omega}+1.5i\Gamma}.
\eea
We write $\tilde{\bf e}(0)=\tilde{\bf e}(0-)-iV{\boldsymbol \rho}$ and find ${\boldsymbol \rho}=\f{3\sqrt{2}V(e_3(k_1)+e_3(k_2))}{\sqrt{\pi}(E_{\bf k}-2\tilde{\Omega}+2i\Gamma)}$. Now we can calculate $\tilde{\bf e}(x>0)$ from Eq.\ref{33tse2}
\bea
\tilde{\bf e}(x)&=&\mathcal{C}_3 e^{i(E_{\bf k}-\Omega+\f{3}{2}i\Gamma)x}+\f{3}{\sqrt{2\pi}}(\tau_3(k_2)e_3(k_1)e^{ik_2x}\nn\\&+&\tau_3(k_1)e_3(k_2)e^{ik_1x}),\\
\mathcal{C}_3&=&18iV e_3(k_1)e_3(k_2)-\f{6\sqrt{2}i\Gamma}{\sqrt{\pi}}\f{e_3(k_1)+e_3(k_2)}{E_{\bf k}-2\tilde{\Omega}+2i\Gamma}.\nn
\eea
\section{Two-photon scattering by four two-level atoms}
\label{4atoms}
The calculation in this appendix is similar to the Appendix \ref{3atoms}, and it can be carried out for many atoms. The Hamiltonian of four two-level atoms coupled to a tightly focused weak light beam (or the atom-photon interaction in a one-dimensional waveguide) can be written after an even-odd transformation for the photon modes as, $\mathcal{H}=\mathcal{H}_e+\mathcal{H}_o$  where 
\bea
\mathcal{H}_e+\mathcal{H}_o&=&-i v_g\int dx(a^{\dagger}_{e}(x)\partial_xa_{e}(x)+a^{\dagger}_{o}(x)\partial_xa_{o}(x))\nn\\&+&\sum_{l=1}^{4}\big[V(a_e^{\dg}(0)\sigma_{l-}+H.c.)+\tilde{\Omega}_la^{\dg}_{el}a_{el}\big].
\eea
Now we set again $v_g=1$. We write a two-photon scattering state for two incident photons with energy, $E_{\bf k}=k_1+k_2$ as
\begin{widetext}
\bea
|k_1,k_2\ra&=&\int dx_1dx_2\Big[A_2\big\{ \mathrm{g}(x_1,x_2)\f{1}{\sqrt{2}}a^{\dg}_e(x_1)a^{\dg}_e(x_2)+\sum_{l=1}^4\mathrm{e}_l(x_1)\delta(x_2)a^{\dg}_e(x_1)\sigma_{l+}+\sum_{m>l=1}^4\mathrm{e}_{lm}\delta(x_1)\delta(x_2)\sigma_{l+}\sigma_{m+}\big\}\nn\\&+&B_2\big\{\mathrm{j}(x_1;x_2)a^{\dg}_e(x_1)a^{\dg}_o(x_2)+\sum_{l=1}^4\mathrm{f}_l(x_2)\delta(x_1)a^{\dg}_o(x_2)\sigma_{l+}\big\}+C_2\mathrm{h}(x_1,x_2)\f{1}{\sqrt{2}}a^{\dg}_o(x_1)a^{\dg}_o(x_2)\Big]|\varnothing \ra.
\label{4wavefn}
\eea
\end{widetext}
Now we can write seventeen coupled linear-differential equations from the two-photon Sch{\"o}dinger equation to find the seventeen unknown amplitudes in the two-photon scattering state in Eq.\ref{4wavefn}. This is similar to the three atoms. We simplify the problem here again by choosing identical atoms, $\Omega_l=\Omega, ~\gamma_l=\gamma$ for $l=1,2,3,4$. 

 Next we define for the identical atoms, $\mathrm{g}(x_1,x_2)\equiv g_4(x_1,x_2),~\mathrm{e}_1(x)+\mathrm{e}_2(x)+\mathrm{e}_3(x)+\mathrm{e}_4(x)=\tilde{\mathrm{e}}(x)$ and $\mathrm{e}_{12}+\mathrm{e}_{13}+\mathrm{e}_{14}+\mathrm{e}_{23}+\mathrm{e}_{24}+\mathrm{e}_{34}=\mathrm{\rho}$. We find from the two-photon Schr{\"o}dinger equation, 
\bea
&&\Big(-i\partial_{x_1}-i\partial_{x_2}-E_{\bf k}\Big)g_4(x_1,x_2)+\f{V}{\sqrt{2}}\tilde{\mathrm{e}}(x)=0, \label{44tse1} \\
&&\Big(-i\partial_x-E_{\bf k}+\tilde{\Omega}\Big)\tilde{\mathrm{e}}(x)+\f{4V}{\sqrt{2}}[g_4(0,x)+g_4(x,0)]\nn\\&&+2V\mathrm{\rho}\delta(x)=0, \label{44tse2} \\
&&(2\tilde{\Omega}-E_{\bf k})\mathrm{\rho}+3V\tilde{\mathrm{e}}(0)=0.\label{44tse3}
\eea
We get from the last equation $\mathrm{\rho}=3V\tilde{\mathrm{e}}(0)/(E_{\bf k}-2\tilde{\Omega})$. We find the following discontinuity relations from these above relations, $g_4(x,0+)=g_4(x,0-)-\f{iV}{\sqrt{2}}\tilde{\mathrm{e}}(x)$ and $\tilde{\mathrm{e}}(0+)=\tilde{\mathrm{e}}(0-)-2iV\mathrm{\rho}$. Now we can calculate $\tilde{\mathrm{e}}(x<0)$ from Eq.\ref{44tse2},
\bea
\tilde{\mathrm{e}}(x)&=&\f{2\sqrt{2}}{\sqrt{\pi}}(e_4(k_2)e^{ik_1x}+e_4(k_1)e^{ik_2x}),~{\rm where}\nn\\e_4(k)&=&\f{V}{\sqrt{2\pi}(E_k-\tilde{\Omega}+2i\Gamma)}.
\eea
Next we derive $g_4(x_1,x_2)$ in the region $x_1<0,x_2>0$,
\bea
g_4(x_1,x_2)&=&\f{1}{2\pi\sqrt{2}}(e^{ik_1x_1+ik_2x_2}\tau_4(k_2)\nn\\&+&e^{ik_2x_1+ik_1x_2}\tau_4(k_1)),\label{g4}\\\tau_4(k)&=&1-4iVe_4(k)=\f{E_k-\tilde{\Omega}-2i\Gamma}{E_k-\tilde{\Omega}+2i\Gamma}.\nn
\eea
We write $\tilde{\mathrm{e}}(0)=\tilde{\mathrm{e}}(0-)-iV\mathrm{\rho}$ and $\mathrm{\rho}=\f{6\sqrt{2}V(e_4(k_1)+e_4(k_2))}{\sqrt{\pi}(E_{\bf k}-2\tilde{\Omega}+3i\Gamma)}$. Next we derive $\tilde{\mathrm{e}}(x>0)$ from Eq.\ref{44tse2} using Eq.\ref{g4}, 
\bea
\tilde{\mathrm{e}}(x)&=&\mathcal{C}_4 e^{i(E_{\bf k}-\tilde{\Omega}+2i\Gamma)x}+\f{2\sqrt{2}}{\sqrt{\pi}}(\tau_4(k_2)e_4(k_1)e^{ik_2x}\nn\\&+&\tau_4(k_1)e_4(k_2)e^{ik_1x}),\\
\mathcal{C}_4&=&32iVe_4(k_1)e_4(k_2)-\f{12\sqrt{2}i\Gamma}{\sqrt{\pi}}\f{e_4(k_1)+e_4(k_2)}{E_{\bf k}-2\tilde{\Omega}+3i\Gamma}.
\eea

\end{document}